Craig Jackson
Center for Applied Cybersecurity Research
Indiana University-Bloomington
scjackso@indiana.edu

Robert Templeman
Naval Surface Warfare Center
Crane Division
robert.templeman@navy.mil

# The Third Offset and a Fifth Domain?

## Balancing Game-Changing Innovation and Cyber Risk Mitigation

## ABSTRACT

Cyber has changed the scope of the Navy's mission and is placing new strains on our diplomatic, warfighting, legal, and economic/budgetary processes. Cybersecurity processes and techniques are increasingly critical to our warfighting missions, but they can also inhibit the pace and potential for high impact, game-changing innovation. Throughout its history, the Navy has shown the ability of innovation (in policy, process, and technology) to change the game when our security is on the line. We believe the Navy is capable of dramatically impacting not only the U.S. capabilities in cyber conflict and information operations, but also in cyber defense and information assurance, as well as cybersecurity for our society. While cyber risk management is challenging, the Navy's transition from DoD Information Assurance Certification and Accreditation Process (DIACAP) to the Risk Management Framework (RMF) has the potential to harmonize our cybersecurity efforts with our need (and demonstrated ability to provide) for game-changing strategies, tactics, and technologies. We offer a foundation for the foregoing assertions and recommendations on ways to encourage innovation in the context of effective cyber risk management.

## INTRODUCTION

Information technology has given us a new warfighting domain replete with complexity and attacker advantages. Global players are dedicating enormous resources to wage real war in a virtual battlespace. The United States Navy, while having a heritage dominated by fighting on the world's seas, also engages in the domains of land, air, and space. In recent years, considerable focus has turned to the cyber domain, 'the fifth domain'[1]. There is increasing acknowledgement that attackers presently enjoy a disproportionate advantage in the cyber domain in spite of our apparent resource superiority[2]. We must innovate both our processes and technologies in order to overcome the asymmetries[3]. We live in a precipitous time where we are still learning about the implications that come from a force that leverages technology in a highly interconnected way.

---

[1] "War in the Fifth Domain", The Economist, July 1, 2010, Available at http://www.economist.com/node/16478792.

[2] For a discussion of how "weak" adversaries can establish an asymmetrical advantage and importance of strategy, *see*, Ivan Arreguin-Toft, How the Weak Win Wars: A Theory of Asymmetric Conflict, International Security, Vol. 26, No. 1 (Summer 2001), pp. 93-128.

[3] Oehmen CS, Multari NJ. 2014. "Report on the First Meeting on Asymmetry in Resilience for Complex Cyber Systems"*Asymmetry in Resilience*, September 17-18, 2014, Crystal City, VA. Available at http://cybersecurity.pnnl.gov/publications.stm.

# THE NAVY'S HISTORY OF INNOVATION

The victor in an engagement is generally the adversary who has the military advantage; that is, a decisive asymmetry. Thus, a tacit objective of militaries is to establish and maintain this asymmetry. History shows that innovation is responsible for the greatest improvements of military advantage; our Navy community can take credit for many of these.

Discussions of innovation often focus on novelty and technology, but this scope is too narrow to apply meaningfully to warfighting. Innovation has many dimensions, none of which should be discounted. The Oxford English Dictionary defines innovation in one way as "the introduction of novelties; the alteration of what is established by the introduction of new elements or forms."[4] Game-changing innovations have come in the form of novel or unexpected strategies, tactics, and technologies, including combinations thereof. Innovative technologies and processes have been "game-changing" because they dramatically and quickly rebalance the advantage.

An exemplary case is the US Navy Polaris program, which resulted in the first submarine-launched ballistic missile (SLBM) more than 50 years ago. This complex and risky project produced a number of high-visibility technological advances in a broad range of scientific and engineering disciplines. However, the greatest innovation may have come from advancements in the disciplines of project and risk management. Admiral William Rayborn led the Special Project Office in charge of Polaris and delivered a critical capability *three years ahead of schedule*[5]. The success of this project was attributed to tools still used today, including the Program Evaluation and Review Technique (PERT). The development and use of this management tool revealed the complexities and interdependencies of elements in large projects. Without this process innovation, it is likely that the project would have taken many years longer or been stopped without ever seeing success.

Moreover, the success of the Polaris program teases at one of the critical dimensions of disruptive innovation: A capability like SLBM is not game-changing unless its arrival is timely. Broad spectrum innovation naturally includes advancements in technology, but success may be ultimately determined by the economy of resources, including funding and time.

There have been moments in history where paired strategic and technological innovation have brought such significant advantages that they have been deemed 'offsets'. FitzGerald defines the pursuit of an offset as a strategy that allows one force to overcome or mitigate another force's advantages[6]. This definition describes our position at the beginning of the Cold War: We were at a quantitative disadvantage relative to the Soviet Union and the rest of the Warsaw Pact with regard to the cost and complexity of projecting a conventional force. Leveraging nuclear deterrence offered enormous savings in resources compared to conventional deterrence. By the early 1970s, the Soviet Union had countered our first offset capability and a formal Offset Strategy was employed to tap technology as a force multiplier. The cornerstone of the second offset was the introduction of Global Positioning System (GPS) and precision munitions. Like the first offset, the second offset was effective for a finite period of time. The rest of the world has largely caught up to the United States in technological capability in the forty years since. It is no surprise that in 2014, then-Secretary of Defense Chuck Hagel called for a *third* offset capability[7].

---

[4] http://www.oed.com

[5] Byrne, J.J., "Polaris: Lessons in Risk Management", Multi-Media Publications Inc., 2011.

[6] FitzGerald, B., "Can America Maintain Its Military-Technology Edge?", Available at http://nationalinterest.org/feature/can-america-maintain-its-military-technology-edge-11071.

[7] Hagel, C., "Defense Innovation Days" keynote, September 3rd, 2014, Available at http://www.defense.gov/News/Speeches/Speech-View/Article/605602.

# CYBER AND THE EFFECT ON INNOVATION

Thomas Aquinas captured this maritime truism about risk-taking in his famous 13th century quote, "If the highest aim of a captain were to preserve his ship, he would keep it in port forever.[8]" Shallow inspection reveals the limitation of this strategy. A port berth certainly mitigates some risk, but warships are not effective fighting while moored to a pier and likely find themselves vulnerable to other grave threats.

We believe cyber defense and information assurance efforts can stifle or enable game-changing innovation at a time when innovation is sorely needed.

In DoD, information security has an institutionalized history of being compliance-oriented rather than risk-oriented. Moreover, in many sectors, attempts to move to risk-based approaches to cyber suffer from this compliance-oriented mindset and focus too much on negative risk, rather than on positive risk-taking. Frequent attacks, poor baseline security practices, organizational silos, and the challenges of attribution all compound the problem and lead organizations to an *either/or* approach to cyber: Either we clamp down on risk-taking in dramatic fashion, or we live in denial of the enormity of the problem.

We will not succeed in our mission if we operate at these extremes. To navigate the space where we can operate, while achieving the 'right' level of risk, we have turned to the discipline of risk management.

# THE CHALLENGE OF APPLYING RISK MANAGEMENT TO CYBER

Cyber poses a formidable problem, and risk management tools, as commonly deployed, do not provide a simple fix. Problems can be described by salient properties including their ease of solution. Moreover, problems can be categorized by their difficulty. Most problems that we encounter, including those where risk management approaches are successful, are *tame* problems. A harder class of problems is *wicked* and is characterized by demonstrated resistance to being solved; for these problems, there is no guarantee of solution. Clemente's well-written piece describes how cyber is a wicked problem[9]. Recently, researchers have identified an even harder class of problems: *super wicked.* Levin *et al*. describe these problems as being even more resistant to solution, not because of the problem itself, but because the agent(s) trying to solve it add to the difficulty[10]. We posit that cyber may be a super wicked problem[11].

Consider the two faces of cyber, offense and defense, and the reality that the Navy uses much of the same commercial hardware and software that our adversaries do. Addressing a known vulnerability in our systems may also have the effect of denying us an offensive tool. Our actions clearly have the potential to increase the difficulty of the cyber problem.

This extreme difficulty explains the frustration that we have when trying to apply standard risk management approaches to cyber. Cyber is simply not as tractable as problems like reliability or quality[12]. While there are likely others, we address three specific difficulties of applying standard risk management tools to cyber.

---

[8] Aquinas, T., "Summa Theologica", 1917.

[9] Clemente, D., "International Security: Cyber Security As A Wicked Problem", Chatham House, October 2011, Available at https://www.chathamhouse.org/node/8489#sthash.8YcMcefq.dpuf.

[10] Levin, K., Cashore, B., Bernstein S., and Auld, G., "Overcoming the tragedy of super wicked problems: constraining our future selves to ameliorate global climate change", Policy Sciences, 2012, Available at http://link.springer.com/article/10.1007%2Fs11077-012-9151-0.

[11] We anticipate publishing a separate paper that expands this argument in the near future.

[12] It is important to note that problems like reliability and quality are not independent with cyber. The addition of cyber to the design tradespace must be acknowledged.

The first two difficulties arise from approaches that equate risk management with risk assessment. Risk assessment is not the same as risk management. Risk assessments are one of the tools used in risk management, and they are particularly challenging in the cyber domain. Risk management is fundamentally about informed, coordinated decision-making, and risk assessments are only one component of supporting decision making. Others include established best practices, and strategic goal setting for information security programs. Too much focus on risk assessment is a mistake, particularly where it is at the expense of valuable resources (like time), benefit assessment, and timely action-taking.

In cyber, the first major obstacle to effective risk assessment involves the enumeration of risks. Standard risk management approaches call for listing all known risks. A program manager may claim that that they have enumerated all risks; a counter-argument to this program manager is the problem of `zero-day vulnerabilities'. We can predict the general existence of these vulnerabilities in our systems, but we cannot mitigate them until details of their existence are known.

The second difficulty of conducting cyber risk assessments lies in the assignment of the likelihood estimations to adverse events. When faced with managing the risk posed by the reliability of electronics, there is a corpus of data, and reliable statistical models can be used to measure uncertainty. Within cyber, we face a severe lack of actuarial data or other efficient methods to determine the likelihood of adverse events.

The third difficulty involves the limitations of risk management approaches to non-static boundary conditions. The sinking of the USS Thresher in 1963 led to the introduction of the US Navy SUBSAFE program. The threat addressed by SUBSAFE is pressurized water that may impinge on the integrity of a submarine. The difficulty of addressing this threat is bounded by the invariant laws of physics: Water behaves the same way today as it behaved in 1963. In the cyber domain, attackers evolve intelligently and quickly, and the complexity and interconnectedness of our information systems, and number of vulnerabilities, may be evolving at an even faster pace. While the threats and terrain shift, we understand relatively little about cyber's immutable laws and principles, assuming they exist. Cyber risks are tied to potential adverse occurrences that may not be predicted by statistical or probabilistic models; rather, these risks are under the direct influence of intelligent, persistent, and well-resourced adversaries. Traditional risk management approaches fall short.

## BALANCING INNOVATION AND RISK IN THE CYBER DOMAIN

Cyber risk management is challenging, and in itself does not provide a perfect set of tools for mission assurance. However, when contextualized in an organizational, mission-oriented environment it can be a flexible, cross-organization approach to managing risk and maximizing mission success.

The practical and pragmatic core of any risk management approach to information security is the ability of organizations to make systematic, informed decisions about mitigating, transferring, avoiding, and – perhaps most importantly – accepting risks. Healthy approaches to risk management are aligned closely to enabling the organizational mission. Risk acceptance is particularly critical in any domain where our ability to mitigate, transfer, or avoid risks is limited (*i.e.*, residual risk cannot be reduced to zero), or risk-taking is critical to the organization's mission. Warfighting is a risk-taking process, and innovation follows suit. Risk *management* holds promise for the cyber domain precisely because (when truly embraced) its processes and outcomes acknowledge that we have a mission to accomplish in spite of the risks over which we cannot exert perfect control. The NIST Risk Management Framework (RMF)[13] holds promise for the Navy because we must intentionally accept some cybersecurity risks in order to innovate and achieve our mission.

---

[13] *See*, NIST Special Publications 800-37 Revision 1 and 800-39. Available at http://csrc.nist.gov/publications/PubsSPs.html.

If risk management approaches are treated as compliance, 'check the box' regimes, they lose their utility.

Information security risk can be managed to enable and encourage the kinds of game-changing innovations that have fueled our military superiority. But, this requires a rich understanding of how innovation and risk relate. Just as we roughly quantify the risk level posed by an adverse threat event as approximately equal to the consequence should the event occur times the likelihood of it occurring (*i.e.*, $risk \approx consequence \cdot likelihood$)[14], we make (sometimes explicitly, sometimes implicitly) a similar calculation when we aim for game-changing innovations. When gauging the value of investment in innovation, indeed the value inherent in accepting certain risks, we multiply the level of benefit to our mission should the investment be successful by the likelihood that it will be successful. Where the potential benefit is very high, the likelihood need not be so great.

On the dark side of risk, we often talk about "black swans.[15]" These are very high impact, very low frequency events that can (and eventually do) happen. Because these risks are critical (even existential) and difficult to measure, risk managers of all kinds obsess over them, and rightly so. To reduce the risk associated with black swans we prepare to reduce their impact when they do happen, and (only with the most existential, most controllable threats) attempt to reduce their probability. On the positive side of risk, game-changing innovations are precisely the flip side of the coin. They are very high impact, low frequency changes in processes and technologies that drastically increase our security, resilience, or ability to overcome an adversary. We lay the groundwork for game-changers by focusing our investments on the most critical problems to increase innovation's impact, and distribute that investment among innovation efforts, knowing that not every seed will sprout, to increase game-changer likelihood.

We can and must continue the Navy's legacy of game-changing process and technological innovation, and cyber must enable, not stifle, that progress. Risk management, when embraced and done in the right way, can harmonize reasonable information security and sound risk-taking[16].

## CONCLUSION

In closing, we offer the following recommendations for managing cyber risk to encourage game-changing innovation:

A. Pair process innovation in cyber with technology innovation across the spectrum of naval capabilities in order to maximize the likelihood of game-changers in or across all warfighting domains.
B. Select and deploy risk management techniques consistent with the Risk Management Framework that support not only informed, but also efficient, timely decision making.
C. Base risk and innovation decisions on their mission impact, and contextualize information security risks in the context of mission assurance. Prioritize cyber risk acceptance that is reasonably likely to reduce higher magnitude mission risks.
D. Work horizontally, across acquisition programs and operational commands, bringing key decision makers to the same table, and ensure that those accepting cyber risk are well-positioned to understand the mission impact.
E. Invest in understanding the cyber terrain, its constant attributes, and evolutionary processes. Research and define how the cyber risk terrain differs across operational and R&D environments.

---

[14] DoD Risk, Issue, and Opportunity Guide (DAU), June 2015, Available at http://www.acq.osd.mil/se/docs/RIO-Guide-Jun2015.pdf.

[15] *See*, Taleb, Nassim Nicholas (April 2007). The Black Swan: The Impact of the Highly Improbable (1st ed.). London: Penguin.

[16] *See, generally*, Johnson, Mark W. (November 2010). Risk management and innovation. Bloomberg Business. Available at http://www.businessweek.com/innovate/content/nov2010/id2010118_752981.htm.

In the *Analects*, Confucius wrote, "Study the past if you would define the future." Our study of the Navy's history of innovation tells us that legacy must continue, even more so now that we face a new domain. We must take cyber risks seriously and reduce them where we can reasonably do so, and we (the authors) spend much of our time advocating for as much. However, the nature of the Navy's mission demands that cyber risk be viewed in light of need for game-changing innovation and the risk-taking required to achieve it.

## ACKNOWLEDGEMENTS

The views expressed herein are the personal opinions of the authors and are not necessarily the official views of Indiana University, or the Department of Defense or any military department thereof.

**Craig Jackson, J.D.,** is Chief Policy Analyst at Indiana University's Center for Applied Cybersecurity Research (CACR), where his research interests include risk management, information security program development and governance, legal and regulatory regimes' impact on information security, and identity management. He leads engagements and authors guidance for the Center for Trustworthy Scientific Cyberinfrastructure (CTSC). He is policy lead of the security team for the DHS-funded Software Assurance Marketplace (SWAMP), and he is part of the DOE-funded XSIM (Extreme Scale Identity Management) project. He is a graduate of the IU Maurer School of Law (J.D.'10) and IU School of Education (M.S.'04). As a member of the Indiana bar, Mr. Jackson has represented government and corporate clients in constitutional and tort claims. His research, design, and project management background includes work at the IU School of Education's Center for Research on Learning and Technology and the Washington University School of Medicine in St. Louis. He is a member of Phi Beta Kappa, and was a Lien Honorary Scholar at Washington University in St. Louis.

**Robert Templeman, Ph.D.,** is an engineer at Naval Surface Warfare Center, Crane Division (NSWC Crane) where he leads efforts related to cyber and systems engineering. He is a graduate of Indiana University (Ph.D. in Computer Science '14), the Rose-Hulman Institute of Technology (M.S. in Engineering Management '05) and Purdue University (B.S. in Electrical Engineering '02). Dr. Templeman's research interests span general security and privacy with an emphasis on applications of machine learning. His recent work has resulted in numerous publications, intellectual property, and media exposure. Dr. Templeman's PlaceRaider work was recognized as "Best of 2012" by MIT's Technology Review and he was a Young Researcher at the Heidelberg Laureate Forum in 2014. He proudly served as an NCO with the 1st Marine Division in Iraq in 2003.